\documentclass{article}
\usepackage{spconf,amsmath,graphicx,hyperref}
\usepackage{amsfonts}   

\title{siDPT: siRNA efficacy Prediction via Debiased Preference-Pair Transformer}
\name{Honggen Zhang, Xiangrui Gao, Lipeng Lai\sthanks{corresponding author}}
\address{XtalPi, Inc}
\begin{document}
%
\maketitle
\begin{abstract}
Small interfering RNA (siRNA) is a short double-stranded RNA molecule (~21–23 nucleotides) with the potential to cure diseases by silencing the function of target genes. Due to its well-understood mechanism, many siRNA-based drugs have been evaluated in clinical trials. However, selecting effective binding regions and designing siRNA sequences requires extensive experimentation, making the process costly. As genomic resources and publicly available siRNA datasets continue to grow, data-driven models can be leveraged to better understand siRNA–mRNA interactions. To fully exploit such data, curating high-quality siRNA datasets is essential to minimize experimental errors and noise.
We propose siDPT: \textbf{si}RNA efficacy Prediction via \textbf{D}ebiased \textbf{P}reference-Pair \textbf{T}ransformer, a framework that constructs a preference-pair dataset and designs an siRNA–mRNA interactive transformer with debiased ranking objectives to improve siRNA inhibition prediction and generalization. We evaluate our approach using two public datasets and one newly collected patent dataset. Our model demonstrates substantial improvement in Pearson correlation and strong performance across other metrics.
The code and data can be found here \url{https://github.com/honggen-zhang/siDPT}.

\end{abstract}
\begin{figure*}[h]
     \centering
         \includegraphics[width=1\textwidth]{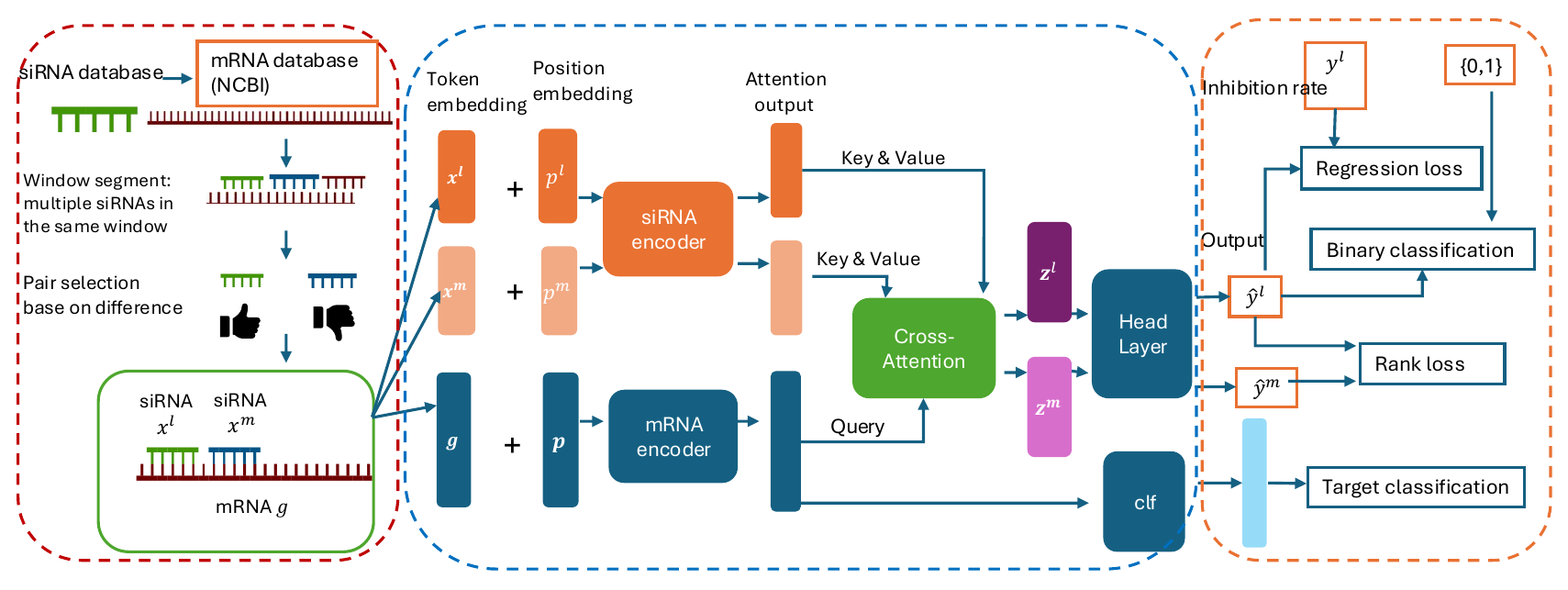}
         \caption{Left: The preference pair data construction. Middle: The siRNA-mRNA interactive transformer. Right: The objective function. }
        \label{fig:diagram}
\end{figure*}
\section{Introduction}
\label{sec:intro}

RNA interference (RNAi) has emerged as a major therapeutic strategy owing to its high specificity and precise gene-targeting capability. Among RNAi approaches, siRNA-based drugs have already been approved for the treatment of conditions such as hypercholesterolemia and rare genetic disorders by the FDA. siRNAs function by knocking down specific genes, thereby preventing mRNA translation and reducing the expression of target proteins.  However, identifying optimal siRNA sequences remains costly and time-consuming, typically requiring several months to determine their efficacy in vivo.

To predict active siRNAs, some researchers analyzed highly effective sequences to identify common biological features, which were then applied to future predictions~\cite{khvorova2003functional}. However, such feature-based approaches were limited by small datasets and often overlooked potential candidates. With the construction of large siRNA databases~\cite{huesken2005design, katoh2007specific}, data-driven methods demonstrated clear advantages by expanding the feature space for prediction. More recently, the emergence of genomic large language models (LLMs) has enabled approaches such as Oligoformer~\cite{bai2024oligoformer}, which incorporate siRNA embeddings extracted from RNA foundation models. While these algorithms show strong performance on benchmark siRNA databases, the issue of dataset bias has received far less attention. In particular, measurement errors in wet-lab experiments introduce uncertainty into the labels: for instance, inhibition rates of 0.51 versus 0.49 cannot serve as a reliable basis for determining which siRNA is superior.

In this paper, we propose siDPT: \textbf{si}RNA efficacy Prediction via \textbf{D}ebiased  \textbf{P}reference-Pair \textbf{T}ransformer. To fully exploit the information contained in siRNA datasets, we first augment the data through preference-pair construction. Specifically, we query the NCBI database to obtain full target mRNA sequences. Each mRNA sequence $g$ is truncated to a length of 100 nucleotides, within which we identify $k$ candidate siRNAs $x_1, x_2, \ldots, x_k$. From the $\binom{k}{2}$ possible pairs, we construct a high-quality preference-pair set $\mathcal{D} = {(g_i, x_i^l, x_i^m)}_i$ based on differences in measured inhibition rates.
We then input $\mathcal{D}$ into a siRNA–mRNA interactive transformer to jointly learn sequence representations. A cross-attention layer is employed, where the mRNA representation serves as the query and the siRNA representation as the key and value, thereby mimicking the biological interaction mechanism between siRNAs and their targets. The resulting attention outputs are passed through a prediction head to estimate siRNA inhibition rates.
To optimize the model, we employ three objectives. (i)\textbf{ Regression loss:} mean squared error between predicted and observed inhibition rates. (ii) \textbf{Rank loss:} preference probability that siRNA $x^l$ is more effective than $x^m$, corrected with a debiased target distribution. (iii) \textbf{Classification loss:} global discrimination of effective versus ineffective siRNAs. In addition, we introduce a gene classification loss to enhance mRNA embeddings and guide representation learning across different target genes (Fig.~\ref{fig:diagram}).

Additionally, we constructed a new evaluation dataset by collecting siRNA sequences from pharmaceutical company patents. To ensure data reliability, we first filtered the sequences based on Pearson correlation across different concentrations measured over 24 hours, retaining only those with correlations greater than 0.80. This process yielded three target-specific datasets. The resulting dataset provides a less noisy benchmark for evaluating siRNA prediction methods. We applied our siDPT on two public datasets and the created patent data. Our experimental results demonstrate that our methods outperform current state-of-the-art methods on several metrics.

\section{Method}

\subsection{Problem Statement}
Let an siRNA sequence be denoted as an antisense strand $x = AUUUCCGG\ldots$. Given a dataset of $N$ siRNAs $\mathcal{X} = {x^1, x^2, \ldots, x^N}$ and their corresponding inhibition rates $\mathcal{Y} = {y^1, y^2, \ldots, y^N}$, our goal is to learn an siRNA encoder $\boldsymbol{x}^i = \boldsymbol{E}_{\text{siRNA}}(x^i)$ and a parameterized regression model $\mathcal{H}$ that predicts $y^i$ from $x^i$:
\begin{equation}\small
\mathcal{L}_{\textbf{MSE}} = \sum_{i=1}^N \|\mathcal{H}(\boldsymbol{x}^i) - y^i\|^2_2,
\end{equation}
where $\|\cdot\|_2$ is the L2 norm. The encoder and regression model are trained by minimizing $\mathcal{L}_{\textbf{MSE}}$.

The non-target area could also affect the pairing ability of siRNA due to mRNA folding into stems or loops in secondary structures. To incorporate this information, we extract a local mRNA segment $g^i$ around the binding site, yielding a pair $(g^i, x^i)$. The mRNA encoder $\boldsymbol{g}^i = \boldsymbol{E}_{\text{mRNA}}(g^i)$ is then combined with the siRNA embedding $\boldsymbol{x}_i$ to form the input to the regression model.

Direct regression can overfit noisy experimental measurements. In practice, the goal is often to select the most effective siRNA among candidates rather than predict exact inhibition rates. Inspired by preference-learning approaches in LLMs~\cite{rafailov2024direct}, we construct preference pairs to learn relative rankings. For a given mRNA segment $g$, let $k$ candidate siRNAs be $x^1, x^2, \ldots, x^k$. We thus form preference pairs $(g, x^l, x^m)$, where $x^l$ is preferred over $x^m$. The Bradley–Terry model~\cite{bradley1952rank} is then used to learn the probability that $x^l$ is more effective than $x^m$ given $g$.

\begin{equation}\small\label{eq:rank}
    p(x^l\succ x^m|g) = \sigma(s_{\theta}(g,x^m)-s_{\theta}(g,x^l))
\end{equation}
where $\sigma(\cdot)$ is the sigmoid function, and $s_{\theta}(\cdot,\cdot)$ is the score function, which is explained as the log-odds of siRNA $x^l$ or $x^m$ being effective for target gene $g$. i.e,
\begin{equation}\small
    s_{\theta}(g,x^l) = \log (\frac{p(x^l \text{  is effective }|g)}{1-p(x^l \text{  is effective }|g)})
\end{equation}
The difference between the score function 
\begin{equation}\small
    s_{\theta}(g,x^l) - s_{\theta}(g,x^m) = \log (\frac{p(x^l\succ x^m|g)}{1-p(x^l\succ x^m|g)})
\end{equation}
 
 Thus, we could obtain the objective function 
\begin{equation}\small
    \min_{\theta} \mathbb{E}_{(g,x^l,x^m) \sim \mathcal{D}} \big[ p(x^l\succ x^m|g)\big]
\end{equation}

However, learning siRNA preference presents three key challenges:
1) Constructing high-quality preference pairs from noisy public datasets.
2) Learning effective representations of siRNAs and target mRNAs to achieve strong predictive performance.
3) Handling varying levels of data noise across target datasets caused by different experimental conditions.

\subsection{SIDPT }
\label{sec:Method}

\subsubsection{Preference Data Construction}
To construct high-quality siRNA preference pairs, we follow a multi-step procedure:

\textbf{Define the binding window:} For each siRNA, we extend the target region to 100 bp—approximately five times the siRNA length—by adding 30 bp upstream and 51 bp downstream of the binding site.
\textbf{Retrieve candidate siRNAs:} Within this window $g$, we identify $k$ siRNAs from the database that also target the region.
\textbf{Build base preference pairs:} We sort the $k$ siRNAs by true inhibition rate in descending order $x_1, x_2, \ldots, x_k$, then construct base pairs by sliding window: $\mathcal{Q} = \{(g, x_1, x_2), \ldots, (g, x_{k-1}, x_k)\}$.
This ensures that each siRNA contributes to training the encoder.
\textbf{Extend with high-confidence pairs:} From all $\binom{k}{2}$ siRNA combinations, we compute the inhibition difference $e = y_i - y_j$. Pairs with $|e| > c$ are considered high-confidence and added to $\mathcal{Q}$, producing $\mathcal{Q}^+$. Where $c$ is the threshold to filter the data. Repeating this process for all siRNAs yields the final high-quality preference dataset: $\mathcal{D} = \bigcup \mathcal{Q}^+.$

\subsubsection{Representation Extraction from Debiased  Preference-Pair Transformer}
Similar to the siRNA silencing mechanism, where the RNA-Induced Silencing Complex (RISC) carries siRNA to recognize and silence complementary mRNA, we design a model to extract representations of siRNA–mRNA binding sites. As shown in Fig.~\ref{fig:diagram}, both siRNA and the corresponding mRNA binding site are tokenized into individual nucleotides ${A, U, G, C}$. Learnable positional embeddings are used to mitigate the effect of repeated tokens in sequences.
The siRNA encoder and mRNA binding site encoder are transformer-based, with 4 heads and 2 layers for siRNA, and 4 heads and 4 layers for the mRNA binding site. The siRNA outputs serve as key and value, while the mRNA outputs serve as query in a cross-attention layer to capture interaction patterns. Following a fully connected layer, we obtain two final outputs, $\boldsymbol{z}^l$ and $\boldsymbol{z}^m$, corresponding to the preferred siRNA $x^l$ and dispreferred siRNA $x^m$. To generate the sequence-level representation for the siRNA–mRNA binding site, we average $\boldsymbol{z}$ across the sequence length and concatenate it with the [CLS] token embedding.
\begin{align}\small
    \boldsymbol{v}_{\text{avg}} &= \frac{1}{L}\sum_{l=1}^L\boldsymbol{z}[l,:]\\
    \boldsymbol{v}_{\text{cls}} &= \boldsymbol{z}[0,:]\\
    s_{\theta}(g,x) &= \mathcal{H}([\boldsymbol{v}_{\text{avg}};\boldsymbol{v}_{\text{cls}}])
\end{align}
where $\mathcal{H}$ plays as a regression to get the prediction inhibition rate.

We also introduce a classifier to assign mRNA binding sites to target gene labels when multiple targets are present in the training set. The corresponding classification loss is:
\begin{equation}\small
\mathcal{L}_{\text{gene\_clf}} = \sum_{(l_g, g)} \log p(l_g \mid \boldsymbol{g}),
\end{equation}
where $l_g$ denotes the gene label of binding site $g$.

\subsubsection{Combine the Global Classification and Debiased Local Rank }
We modify Eq.~\ref{eq:rank} to account for bias in small inhibition differences
$d(x^l, x^m) = |y^l - y^m|$, which may be unreliable. We weight each pair with a noise-aware target distribution $q^{\star}(x^l \succ x^m \mid g)$ using the inhibition difference and a temperature $\beta$:
\begin{equation}\small
    \mathcal{L}_{\text{rank}} = \sum_{(g,x^l,x^m)} q^{\star}(x^l\succ x^m|g)\log p_{\theta}(x^l\succ x^m|g) 
\end{equation}
where 
\begin{equation}\small
    q^{\star}(x^l\succ x^m|g) = \frac{\text{exp}(d_{l,m}/\beta(g))}{\sum_{(a,b)} \text{exp}(\text{exp}(d_{a,b}/\beta(g))}
\end{equation}
Here, $d$ reflects the inhibition difference measured in wet-lab experiments, and $\beta(g)$ encodes label reliability for target $g$: higher $\beta$ smooths $q^\star$ for noisy genes, while lower $\beta$ sharpens it for reliable genes.

We also include a binary classification loss to differentiate positive ($r=1$) and negative ($r=0$) siRNAs: 
\begin{equation}\small
    \mathcal{L}_{\text{binary\_clf}} = \sum_{(r^l,r^m,g,x^l,x^m)} \big [\log p (r^l|s_{\theta}(g,x^l))+ \log p (r^m|s_{\theta}(g,x^m))\big ]
\end{equation}
Additionally, a regression MSE loss provides a numerical constraint:
\begin{equation}\small
    \mathcal{L}_{\text{MSE}} = \sum_{(g,x^l,x^m)} \|\boldsymbol{s}_{\theta}(g,{x}^l)- y^l\|^2_2 +\|\boldsymbol{s}_{\theta}(g,{x}^m)- y^m\|^2_2
\end{equation}
Thus, the final loss function
\begin{equation}\small
    \mathcal{L} = \alpha_1 \mathcal{L}_{\text{MSE}} + \alpha_2 \mathcal{L}_{\text{rank}} + \alpha_3 \mathcal{L}_{\text{binary\_clf}}++ \alpha_4 \mathcal{L}_{\text{gene\_clf}}
\end{equation}
where $\alpha_1, \alpha_2, \alpha_3, \alpha_4$ are hyperparameters used to assign weights to each objective function.
\begin{table*}
  \centering
    \caption{siRNA inhibition prediction results on public datasets. Boldface indicates the best performance.}
  \begin{tabular}{cccc ccc }
    \hline
     Method& \multicolumn{3}{c}{Huesken Dataset} & \multicolumn{3}{c}{Takayuki Dataset}\\
     \cline{2-4} \cline{5-7}
     & AUC & F1 & Pearson & AUC & F1 & Pearson \\ 
    \cline{1-4} \cline{5-7}
    Biopredsi~\cite{huesken2005design} & 0.8664 & 0.8287   & 0.6590 & 0.7576  &0.4379& 0.5287 \\ 
    iScore~\cite{ichihara2007thermodynamic} & 0.8625  & 0.8137  & 0.6538  & 0.7695 &0.0757& 0.5317  \\ 
    DSIR~\cite{vert2006accurate} & 0.8434 & 0.7165  & 0.6272 & 0.7702  &0.5422& 0.5815  \\ 
    Monopoli-RF~\cite{monopoli2023asymmetric} & 0.805  & 0.7276  & 0.5731  & 0.7756 &0.0909& 0.5578 \\ 
    OligoFormer~\cite{bai2024oligoformer} & 0.8725  & 0.8123  & 0.6688 & \textbf{0.8628} &0.5769& 0.6596 \\ 
    siDPT (Ours) & \textbf{0.8873}  &\textbf{0.8339}  & \textbf{0.6741} &0.8519  &\textbf{ 0.6096}& \textbf{0.6624} \\ \hline
  \end{tabular}
  \label{tab:result1}
\end{table*}

\begin{table*}
  \centering
  \caption{siRNA inhibition prediction results on the new patent dataset. Boldface indicates the best performance.}
  \begin{tabular}{c ccc ccc ccc }
    \hline
     Method& \multicolumn{3}{c}{KHK} & \multicolumn{3}{c}{CTNNB1}& \multicolumn{3}{c}{TMPRSS6} \\
     \cline{2-4} \cline{5-7}\cline{8-10}
     & AUC & F1 & Pearson & AUC & F1 & Pearson & AUC & F1 & Pearson \\ 
    \hline
    Biopredsi~\cite{huesken2005design} & 0.5162 & 0.4658  & -0.1831  & 0.5786 & 0.5081 & 0.1924 &0.5289& 0.4059&0.0570 \\ 
    iScore~\cite{ichihara2007thermodynamic} & 0.5278  & 0.500  & -0.1515  & 0.5698 & 0.6531 & 0.1642 &0.5052& 0.3272&0.0100  \\ 
    DSIR~\cite{vert2006accurate} & 0.5828 & 0.4706  & -0.0328  & 0.5770 & 0.6351 & 0.1891 &0.5348& 0.400 & 0.030  \\ 
    OligoFormer~\cite{bai2024oligoformer} & 0.700  & \textbf{0.5833}  & 0.2081  & 0.5396 & 0.6809 & 0.0191&0.5951& 0 & -0.0936 \\ 
    siDPT (Ours) & \textbf{0.8251}  &0.4944  &\textbf{ 0.4967}  & \textbf{0.5948} &\textbf{0.7537} & \textbf{0.1946} &\textbf{ 0.7338}& \textbf{0.4444} &\textbf{ 0.4149} \\ \hline
  \end{tabular}
  \label{tab:Patent}
\end{table*}

\section{EXPERIMENT}
\label{sec:majhead}
\subsection{Dataset}
In this study, we use two public datasets: Huesken~\cite{huesken2005design} (29 targets, 2,431 siRNAs) and Takayuki~\cite{katoh2007specific} (1 target, 702 siRNAs). We also constructed a new dataset from patent documents. From this dataset, three targets—KHK, CTNNB1, and TMPRSS6—were selected based on high-confidence inhibition measurements, with Pearson correlation across different concentrations exceeding 0.80. To ensure comparability, all experiments were restricted to the same cell line (HEP3B) and a uniform 24-hour treatment. This yielded 248 siRNAs targeting CTNNB1, 212 targeting TMPRSS6, and 72 targeting KHK.
\vspace{-0.3cm}
\subsection{Main result: Inhibition Prediction}
\label{ssec:subhead}
\subsubsection{Evaluation on the Public dataset}
To evaluate knockdown efficacy in vivo, we compare our method with five siRNA inhibition prediction tools: Biopredsi~\cite{huesken2005design}, i-Score~\cite{ichihara2007thermodynamic}, DSIR~\cite{vert2006accurate}, Monopoli-RF~\cite{monopoli2023asymmetric}, and OligoFormer~\cite{bai2024oligoformer}. For the Huesken dataset, we follow the training/test split from Biopredsi~\cite{huesken2005design}. For the Takayuki dataset, results are averaged over five five-fold cross-validations. Predictions are evaluated using ROC-AUC, F1 score, and Pearson correlation. Biopredsi, i-Score, and DSIR results are taken from the literature~\cite{ichihara2007thermodynamic}, while Monopoli-RF and OligoFormer are re-implemented on the respective training sets.

On the Huesken dataset (Table~\ref{tab:result1}), our model outperforms existing tools across all metrics. On the Takayuki dataset, our method achieves the best F1 and Pearson correlation, with AUC slightly below OligoFormer. Notably, our model achieves the highest Pearson correlation on both datasets, which is critical for siRNA selection when positive/negative labels are unavailable.

\subsubsection{Evaluation on the Patent Dataset (Zero-Shot)}
We select the 4 tools which has been trained on the Huesken dataset to apply to the new patent dataset. We are evaluating each of the three target datasets separately using AUC, F1, and Pearson correlation. For the OligoFormer and our model, we train them on the Huesken dataset and then test them on the patent dataset. As shown in the Table~\ref{tab:Patent}, 1)our method achieves the best performance on all three datasets in a zero-shot setting, demonstrating superior generalization to unseen targets and practical utility. 2) The Pearson correlation has been largely improved compared to other methods on KHK and TMPRSS6. It improved from 0.2081 to 0.4976 for KHK and from 0.057 to 0.4119 for TMPRSS6, demonstrating substantially stronger performance than existing siRNA prediction tools.
\vspace{-0.5cm}
\section{Related work}
\label{sec:related}

Early siRNA efficacy prediction relied on handcrafted features, including thermodynamic stability~\cite{naito2009sidirect}, nucleotide composition~\cite{khvorova2003functional}, and positional rules~\cite{katoh2007specific}. With the release of large-scale datasets~\cite{huesken2005design}, machine learning methods became feasible, e.g., i-Score and DSIR (linear regression), SVM-based method~\cite{wang2010predicting}, and ensemble models such as AdaBoost~\cite{monopoli2023asymmetric}.
Deep learning further advanced prediction with neural networks~\cite{han2018sirna, bereczki2025mitigating}, graph neural networks~\cite{la2022graph}, and latent representation learning~\cite{he2017predicting}. Transformer-based models~\cite{liu2024attsioff} and RNA foundation model embeddings (RNA-FM~\cite{chen2022interpretable}, Evo~\cite{brixi2025genome}, mRNA2vec~\cite{zhang2025mrna2vec}, Oligoformer~\cite{bai2024oligoformer}) now represent the state of the art.
However, dataset bias remains underexplored~\cite{long2024sirnadiscovery}, limiting the robustness of existing approaches.
\vspace{-0.5cm}
\section{Conclusion}
In this work, we proposed siDPT, a debiased preference-pair transformer for siRNA inhibition prediction. By constructing high-quality preference pairs, integrating siRNA-mRNA interactive transformer and debiased ranking loss, our model effectively mitigates noise from experimental measurements. Extensive evaluations on public datasets and a newly curated patent dataset demonstrate that siDPT outperforms existing siRNA prediction tools. Notably, our method shows strong zero-shot generalization to unseen targets, making it a practical tool for siRNA selection in drug development.

\vfill\pagebreak

\bibliographystyle{IEEEbib}
\bibliography{ICASSP2026_Paper_Templates/main}

\end{document}